 \documentclass[final,3p,times,twocolumn]{elsarticle}

\usepackage{graphicx}

\usepackage{amssymb}
 \usepackage{amsthm}

\journal{Physics Letters B}

\begin{document}

\begin{frontmatter}


\title{Hypernuclei, dibaryon and antinuclei production in high energy heavy ion collisions: Thermal production vs. Coalescence\tnoteref{label10}}


 \author[label1,label2]{J.~Steinheimer}
 \author[label2,label3]{K.~Gudima} 
  \author[label2,label4]{A.~Botvina} 
  \author[label2,label5]{I.~Mishustin} 
  \author[label2]{M.~Bleicher} 
  \author[label2,label3]{H.~St\"ocker} 
\address[label1]{Lawrence Berkeley National Laboratory, 1 Cyclotron Road, Berkeley, CA 94720, USA}
\address[label2]{FIAS, Johann Wolfgang Goethe University,
Frankfurt am Main, Germany}
\address[label3]{Institute of Applied Physics, Academy of Sciences of Moldova, 
MD-2028 Kishinev, Moldova}
\address[label4]{Institute for Nuclear 
Research, Russian Academy of Sciences, 117312 Moscow, Russia}
\address[label5]{Kurchatov Institute, Russian Research Center,
123182 Moscow, Russia}
\address[label6]{GSI Helmholtzzentrum f\"ur Schwerionenforschung GmbH, Planckstr.~1, D-64291 Darmstadt, Germany}



%

%

%

%

\begin{abstract}
We study the production of (hyper-)nuclei and di-baryons in most central heavy
Ion collisions at energies of $E_{lab}=1-160 A$ GeV. In particular we are
interested in clusters produced from the hot and dense fireball. The formation
rate of strange and non-strange clusters is estimated by assuming thermal
production from the intermediate phase of the UrQMD-hydro hybrid model and
alternatively by the coalescence mechanism from a hadronic cascade model. Both
model types are compared in detail. For most energies we find that both
approaches agree in their predictions for the yields of the clusters. Only for
very low beam energies, and for di-baryons including $\Xi$'s, we observe
considerable differences. We also study the production of anti-matter clusters
up to top RHIC energies and show that the observation of anti-$^4He$ and even
anti-$^4_{\Lambda}He$ is feasible. We have found a considerable qualitative
difference in the energy dependence of the strangeness population factor $R_H$
when comparing the thermal production with the coalescence results.

\end{abstract}


\end{frontmatter}


\section{Introduction}

Relativistic heavy ion collisions are an abundant source of strangeness. As strange 
quarks have to be newly produced during the hot and dense stage of the collision, 
they are thought of carrying information on the properties of the matter that was 
created \cite{Koch:1986ud}. Together with other probes like the elliptic flow and jet quenching, 
the enhancement of strange particle production is discussed \cite{Adams:2005dq,Back:2004je,Arsene:2004fa,Adcox:2004mh,Ollitrault:1992bk,Rischke:1996nq,Sorge:1996pc,Heiselberg:1998es,Scherer:1999qq,Soff:1999yg,Brachmann:1999xt,Csernai:1999nf,Zhang:1999rs,Kolb:2000sd,Bleicher:2000sx,Stoecker:2004qu,Zhu:2005qa,Petersen:2006vm,Gazdzicki:2004ef,Gazdzicki:1998vd} as a possible signal for the creation of a deconfined phase.\\
Although abundantly produced, the strong interactions of strange hadrons are not 
well understood. Such interactions are not only important for the description of 
the hadronic phase of a heavy ion collision but also play an important role for 
the description of dense hadronic matter. In this context hyperon interactions are 
key to understand the phase structure of QCD at large densities and the interior of 
compact stars. One way to tackle the problem of hyperon interactions is to study 
the formation of hyperclusters and/or hypernuclei. Hypernuclear physics offers a 
direct
experimental way to study hyperon--nucleon ($YN$) and hyperon--hyperon
($YY$) interactions ($Y=\Lambda,\Sigma,\Xi,\Omega$). 
The nucleus serves as a laboratory offering the unique opportunity 
to study basic properties of hyperons and their interactions. Even the confirmation 
or exclusion of the existence for such objects can be used as an input for models 
that try to describe hyperonic interactions.\\
More exotic forms of deeply bound objects with strangeness have been proposed 
\cite{Bodmer:1971we} 
as states of matter, either consisting of baryons or quarks.  
The H di-baryon was predicted by Jaffe
\cite{Jaffe:1976yi} and later, many more bound di-baryon states with strangeness were 
proposed
using quark potentials \cite{Goldman:1987ma,Goldman:1998jd} or the Skyrme 
model \cite{Schwesinger:1994vd}.
However, the non-observation of multi-quark bags, e.g. strangelets is still one of 
the open problems of 
intermediate and  high energy physics. Lattice calculations suggest that the 
H-dibaryon is a weakly unbound system \cite{Wetzorke:2002mx}, while recent lattice studies report 
that there could be strange di-baryon systems including $\Xi$'s that can be 
bound \cite{Beane:2011iw}. Because of the size of these clusters lattice studies are usually very 
demanding on computational resources and have large lattice artifacts, it is not clear if Lattice QCD predicts a loosely bound H-dibaryon
or if it is unbound \cite{Beane:2010hg,Beane:2011xf,Inoue:2010es,Buchoff:2012ja}. An experimental confirmation of such a state would therefore be an enormous advance in the 
understanding of the hyperon interaction.\\
For completeness we also include in our analysis a hypothetical N$\Lambda$
di-baryon with mass 2.054 GeV (see Table 1), a weakly bound state of a
$\Lambda$-hyperon and a neutron. The search for such an exotic object is underway
at GSI \cite{Saito}.\\
Hypernuclei are known to exist and be produced in heavy Ion collisions already 
for a long time \cite{nucl-th/9412035,Ahn:2001sx,Takahashi:2001nm,arXiv:1010.2995}. 
The recent discoveries of the first anti-hypertriton \cite{star2010} and anti-$\alpha$ \cite{star2011} (the 
largest anti-particle cluster ever reported) has fueled the interest in the field 
of hypernuclear physics. 
Metastable exotic multi-hypernuclear objects (MEMOs)
as well as purely hyperonic systems of $\Lambda$'s and $\Xi$'s
were introduced in \cite{Schaffner:1992sn,Schaffner:1993nn} as the hadronic 
counterparts to
multi-strange quark bags \cite{Gilson:1993zs,SchaffnerBielich:1996eh}.\\

Hypernuclear clusters can be produced and studied in various experimental setups, e.g. from proton or anti-proton induced reactions \cite{Gaitanos:2011fy} as well as pion and kaon beams \cite{Faessler:1974xn,Chrien:1979wu,Akei:1990gb,Dohrmann:2004xy,Hashimoto:2006aw}.
In this work we will focus on the production of hypernuclei in high energy 
collisions of Au+Au ions \cite{Baltz:1993jh}. In such systems strangeness is produced abundantly 
and is likely to form clusters of different sizes. Our aim is to 
determine which processes are most efficient in searching for hypernuclei including 
exotic ones. Presently, we can discriminate two 
distinct mechanisms for hypercluster formation in heavy ion collisions. First, 
the absorption of hyperons in the spectator fragments of non central heavy ion 
collisions. In this scenario one is interested in hyperons which propagate with 
velocities close to the initial velocities of the nuclei, i.e., in the vicinity 
of nuclear spectators \cite{Ko:1985gp,Gaitanos:2007mm,Gaitanos:2009at,Botvina:2011jt}. 
The hyper-systems obtained here are rather large and moderately excited, decaying into hyperfragments later on \cite{Botvina:2011jt,Botvina:2007pd}.
Alternatively, (hyper-)nuclear clusters can emerge from the hot and dense fireball 
region of the reaction. In this scenario the cluster is formed at, or shortly after, 
the (chemical-)freeze out of the system. A general assumption is, that these 
clusters are then formed through coalescence of different newly produced 
hadrons \cite{Scheibl:1998tk}. To estimate the production yield we can employ two distinct approaches which allow us to estimate the theoretical uncertainties associated with different treatment of the process.
First we use a hadronic transport model to provide us with the phase space information of 
all hadrons produced in a heavy ion collision. This information then serves as an 
input for a coalescence prescription. On the other hand it has been shown \cite{Becattini:1997rv,Cleymans:1990mn,Andronic:2005yp} that thermal models consistently describe the production yields of hadrons (and nuclei 
\cite{Andronic:2008gu}) very well. We can therefore assume thermal production of clusters from a fluid 
dynamical description to heavy ion collisions.\\
Both approaches differ significantly in their assumptions and one would expect 
to obtain different results, depending on the method used. Hence it has been proposed (e.g. see \cite{Cho:2010db,Cho:2011ew}) that the yield of an exotic 
hadronic state may depend strongly on its structure.
The purpose of this paper is therefore to, comprehensively, compare hypernuclei and di-baryon production 
from a coalescence and thermal/hydrodynamical approach, and interpret the differences. One particular important point is that we deliberately compared two distinctively different models to explore the robustness of our predictions. In this way we can estimate systematic differences introduced by the two models features, for example a difference in the baryon stopping or hyperon phase space distributions.

\section{Thermal production from the UrQMD hybrid model}

\begin{table}[t]
		\begin{tabular}{|c|c|c|c|c|}
		\hline 
		Cluster & Mass [GeV] & Chem. Pot. & Spin Deg.\\ \hline\hline
		$d$& 1.878 & $2 \mu_B$ & 3\\ \hline	
		$\{N \Lambda\}$& 2.054 & $2 \mu_B - \mu_S$ & 3\\ \hline			
		$\{\Lambda \Lambda\}$& 2.232 & $2 \mu_B - 2 \mu_S$& 1 \\ \hline
	  $\{N \Xi\}$& 2.260 & $2 \mu_B - 2 \mu_S$& 1 \\ \hline	
		$\{\Lambda \Xi\}$& 2.437 & $2 \mu_B - 3 \mu_S$& 1  \\ \hline	
		$\{\Xi \Xi\}$ & 2.636 & $2 \mu_B - 4 \mu_S$& 1 \\ \hline	
		$He^3$& 2.817 & $3 \mu_B$& 2  \\ \hline
		$He^4$& 3.756 & $4 \mu_B$& 1 \\ \hline
		$^{3}_{\Lambda}H$ & 2.994 & $3 \mu_B -\mu_S$& 2\\ \hline
		$^{4}_{\Lambda}H$ & 3.933 & $4 \mu_B -\mu_S$& 1\\ \hline
		$^{5}_{\Lambda}He$ & 4.866 & $5 \mu_B -\mu_S$& 2\\ \hline
		$^{4}_{\Lambda \Lambda}He$ & 4.110 & $4 \mu_B -2 \mu_S$& 1\\ \hline
		\end{tabular}
	\caption{Properties of all considered multibaryonic states \label{table1}}
\end{table}

The hybrid approach used in this work is based on the integration of a hydrodynamic 
evolution into the UrQMD transport model 
\cite{Petersen:2008dd,Petersen:2008kb,Steinheimer:2007iy}. 
During the first phase of the evolution the particles are described by UrQMD 
as a string/hadronic cascade. Once the two colliding nuclei have passed through 
each other the hydrodynamic evolution starts at the time 
$t_{start}=2R/\sqrt{\gamma_{c.m.}^2-1}$, where $\gamma_{c.m.}$ denotes the Lorentz 
factor of the colliding nuclei in their center of mass frame. 
While the spectators continue to propagate in the cascade, all other particles, 
i.e. their baryon charge densities and energy-momentum densities, are mapped to 
the hydrodynamic grid. By doing so one explicitly forces the system into a local 
thermal equilibrium for each cell. In the hydrodynamic part we solve the 
conservation equations for energy and momentum as well as the net baryon number 
current, while for the net strange number we assume it to be conserved and equal 
to zero locally. Solving only 
the equations for the net baryon number is commonly accepted in hydrodynamical 
models, although we have shown in earlier \cite{Steinheimer:2008hr}
publications that net strangeness may fluctuate locally. It is planned to also 
implement an explicit propagation for the net strange density.\\
The hydrodynamic evolution is performed using the SHASTA algorithm 
\cite{Rischke:1995ir}. At the end of the hydrodynamic phase the fields are 
mapped to particle degrees of freedom using the Cooper-Frye equation 
\cite{Cooper:1974mv} with the properties of the clusters, which serve as input 
for the computation, being listed in Table \ref{table1}. The transition from 
the hydrodynamic prescription to the transport simulation is done gradually in 
transverse slices of thickness 0.2 fm, once all cells in a given slice have an 
energy density lower than five times the ground state energy density (see also 
\cite{Steinheimer:2009nn}). The temperature at $\mu_B=0$ which corresponds to such 
a switching density
is roughly $T=170$ MeV which is close to what is expected to be the critical 
temperature. Detailed information of the transition curve in the phase diagram 
can be found in \cite{Petersen:2008dd}.
In this work we neglected final state interactions of the clusters produced. This can be justified, as previous works have shown that
final state interactions reduce e.g. the deuteron yield by only about $20 \%$ \cite{Oh:2009gx}. \\
For an extensive description of the model the reader is referred to 
\cite{Petersen:2008dd,Steinheimer:2009nn}.\\

\section{Coalescence from the DCM--QGSM approach}

\begin{figure}[t]
\begin{center}
\includegraphics[width=0.5\textwidth]{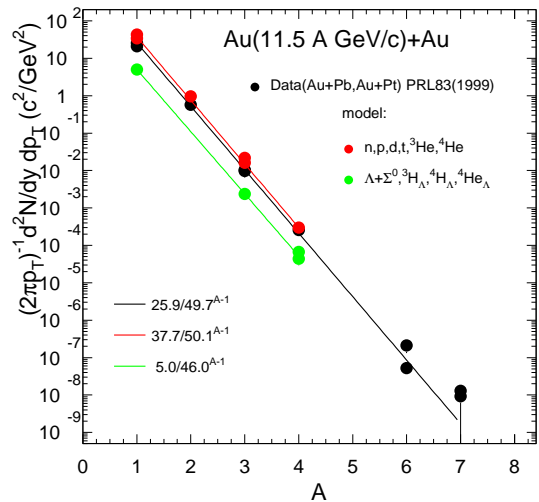}
\caption{Mass dependence of calculated invariant yields of light fragments and 
hyperfragments produced in central Au+Au collisions at 11.5 A GeV/c compared with 
experimental data \cite{Armstrong:2000gz} for Au + Pb collisions. The lines are empirical 
interpolations of the results. 
\label{fig1coal}
}
\end{center}
\end{figure}

Another model used to describe the dynamical stage of the reaction is the 
intra-nuclear cascade model developed in Dubna \cite{Toneev:1990vj,Amelin:1989ve}. 
(We refer to it as the Dubna Cascade Model - DCM.) The DCM is based on 
the Monte-Carlo solution of a set of the 
Boltzmann-Uehling-Uhlenbeck relativistic kinetic equations with 
the collision terms, including cascade-cascade 
interactions. For particle energies below 1~GeV it is sufficient to 
consider only nucleons, pions and deltas. The model includes a proper 
description of pion and baryon dynamics for particle production and 
absorption processes. 
At energies higher than about 10~GeV, the Quark-Gluon String Model (QGSM) 
is used to describe elementary hadron collisions. 
The QGSM considers the two lowest SU(3) multiplets in 
mesonic, baryonic and antibaryonic sectors, so interactions between almost 
70 hadron species are treated on the same footing. 
The above noted two energy extremes were bridged by the QGSM extension 
downward in the beam energy \cite{Amelin:1989ve}. 

For the present study the coalescence model has been modified in comparison 
with its initial formulation in \cite{Toneev:1983cb}. As usual, the coalescence model 
forms a deuteron from a proton and a neutron produced after the cascade stage of 
reaction if their relative momenta are within a sphere of radius $p_C$, comparable 
to the deuteron's momentum. The same momentum criterion can be used to describe 
formation of tritons, $^3$He, and $\alpha$-particles. In particular, the parameters 
$p_C(d)$=90 ,  $p_C(t)$=108 , $p_C(^3$He$)$=108 , and $p_C(\alpha)$=115 (MeV/c) were adopted 
to reproduce the experimental data \cite{Toneev:1983cb}. An approach disregarding the spacial 
coordinates of nucleons can be justified only for collisions with moderate energy 
deposition in nuclei since the region for final state interaction is small enough. 
However, this is not the case for central heavy ion collisions. 
Here we assume that the coalescence criterion 
used to form the composite particles includes the proximity of nucleons both in 
the momentum and coordinate space. The coordinate coalescence parameters are determined 
by the relation $r_C=\hbar / p_C$, with the same values of $p_C$  as were used 
in \cite{Toneev:1983cb}. As a first approximation we use the same coalescence parameters for both 
conventional fragments and hyperfragments. An example of the calculated invariant yields of the fragments 
produced in the central  Au + Au collisions at projectile momentum $11.5 A$ GeV is 
shown in Fig.~\ref{fig1coal}. One can understand that at this energy the coalescence 
model reproduces qualitatively the experimental data for conventional fragments. 
The fragments yields fit very close to exponential dependence with a penalty 
factor of approximately 50 for each nucleon added in agreement with the data. 
Due to the fact that the same coalescence parameters were used  a similar penalty 
factor is obtained for hyperfragments, which is supplemented by additional 
suppression if the neutron is replaced by a $\Lambda$.\\
For the following results we fixed the coalescence parameters as described, with a fit to the data at $11.5 A$ GeV, and assume that they do not change with beam energy. This allows us to predict cluster production over a wide range of experimental setups.

\begin{figure}[t]
\begin{center}
\includegraphics[width=0.5\textwidth]{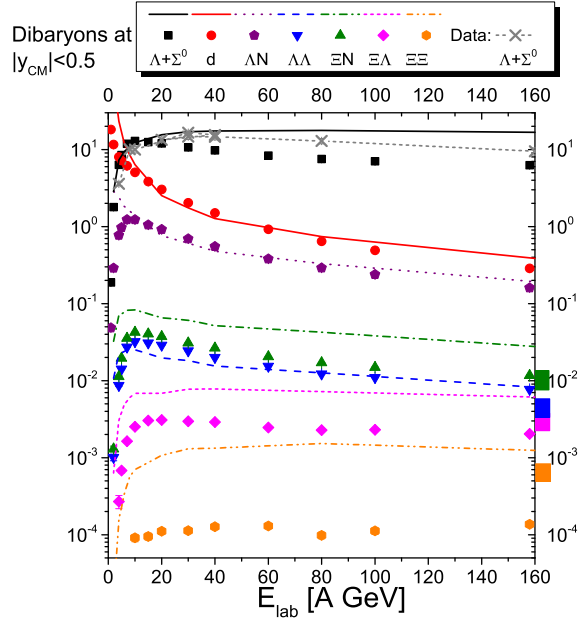}
\caption{Yields per event of different di-baryons in the mid rapidity region ($|y|<0.5$) of most central collisions of Pb+Pb/Au+Au. Shown are the results from the thermal production in the UrQMD hybrid model (lines) as compared to coalescence results with the DCM model (symbols). The small bars on the right hand axis denote results on di-baryon yields from a previous RQMD calculation at $\sqrt{s_{NN}}=200$ GeV \cite{SchaffnerBielich:1999sy}. In addition, the black lines and symbols depict results for the production rate of $\Lambda$'s from both models, compared to data (grey crosses) from \cite{Ahmad:1991nv,Mischke:2002wt,Alt:2008qm}.
\label{dibmidy}
}
\end{center}
\end{figure}

\begin{figure}[t]
\begin{center}
\includegraphics[width=0.5\textwidth]{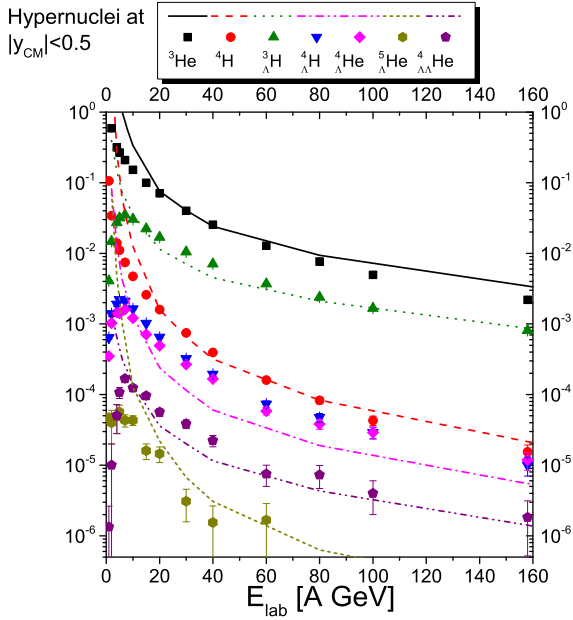}
\caption{Yields per event of different (hyper-)nuclei in the mid rapidity region ($|y|<0.5$) of most central collisions of Pb+Pb/Au+Au. Shown are the results from the thermal production in the UrQMD hybrid model (lines) as compared to coalescence results with the DCM model (symbols). 
\label{hypmidy}
}
\end{center}
\end{figure}

\section{Results}

Figures \ref{dibmidy} and \ref{hypmidy} show our results for the mid rapidity 
yields ($|y|<0.5$) of di-baryons and hypernuclei as a function of the beam energy 
$E_{lab}$. In our calculations we considered most central ($b<3.4$ fm) Pb+Pb/Au+Au 
collisions at $E_{lab}=1$ - $160 A$ GeV. In addition, figure \ref{dibmidy} shows 
the $\Lambda$ yield (black lines and squares) for the two different models compared 
to data \cite{Ahmad:1991nv,Mischke:2002wt,Alt:2008qm}. In these figures, the UrQMD hybrid model calculations are shown as 
lines, while the DCM Coalescence results are depicted as symbols. A striking feature 
of our comparison is that, above $E_{lab} \sim 10 A$ GeV, both computations for most 
(hyper-)nuclei and di-baryons agree very well. At lower energies the 
strange cluster production is suppressed in the transport model due to the non-equilibrium of 
strangeness. In the thermal calculations restrictions of energy and momentum 
conservation, resulting in a phase space reduction for produced strange particles, 
strongly decreases strange particle yields \cite{Becattini:1997rv,Cleymans:1990mn,Andronic:2005yp}. This behavior was also 
observed in a core-corona implementation in the hybrid model \cite{Steinheimer:2011mp}.\\

\begin{figure}[t]
\begin{center}
\includegraphics[width=0.5\textwidth]{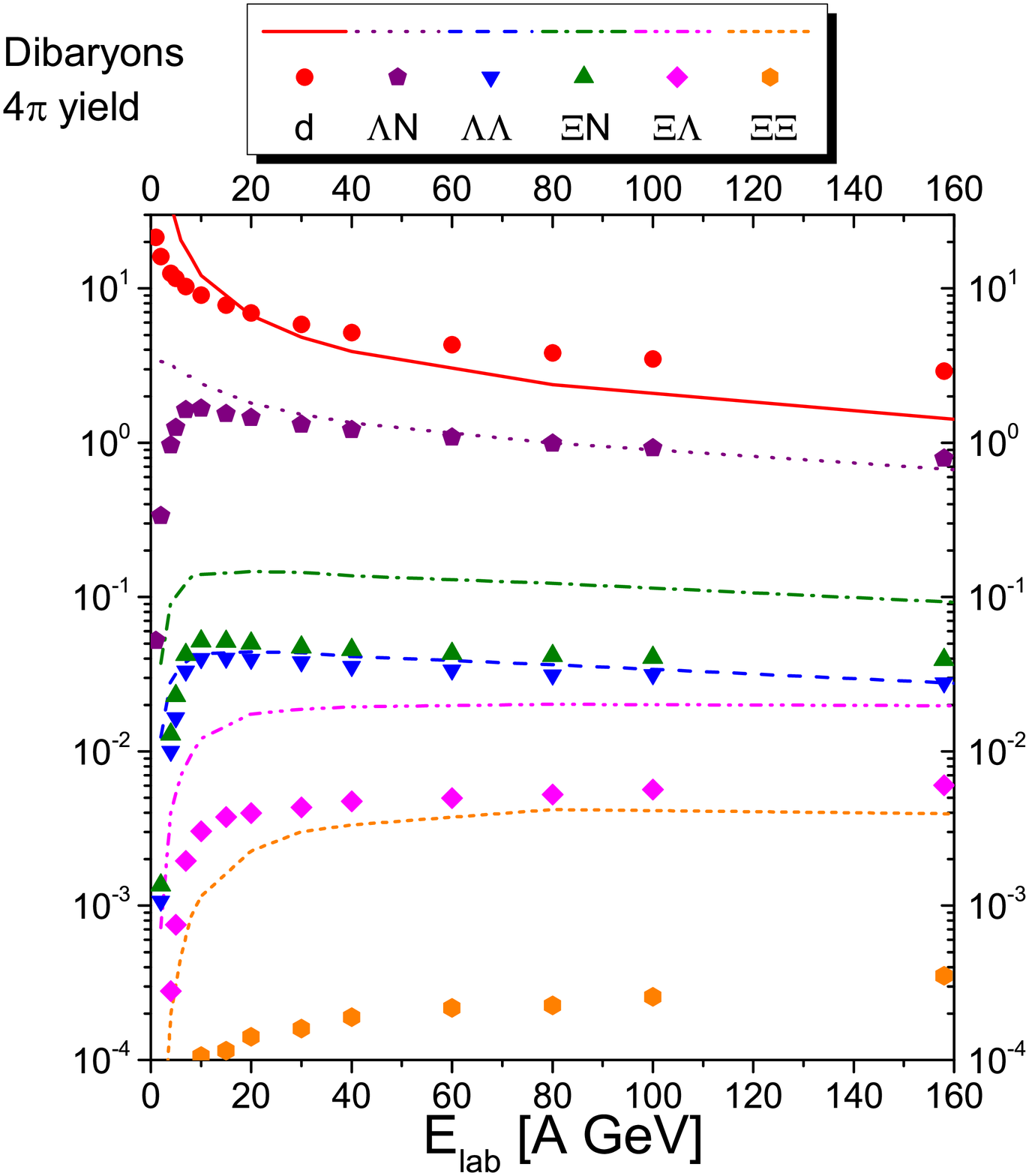}
\caption{Full acceptance yields per event of different di-baryons created in most central collisions of Pb+Pb/Au+Au. Shown are the results from the thermal production in the UrQMD hybrid model (lines) as compared to coalescence results with the DCM model (symbols). 
\label{dibally}
}
\end{center}
\end{figure}

\begin{figure}[t]
\begin{center}
\includegraphics[width=0.5\textwidth]{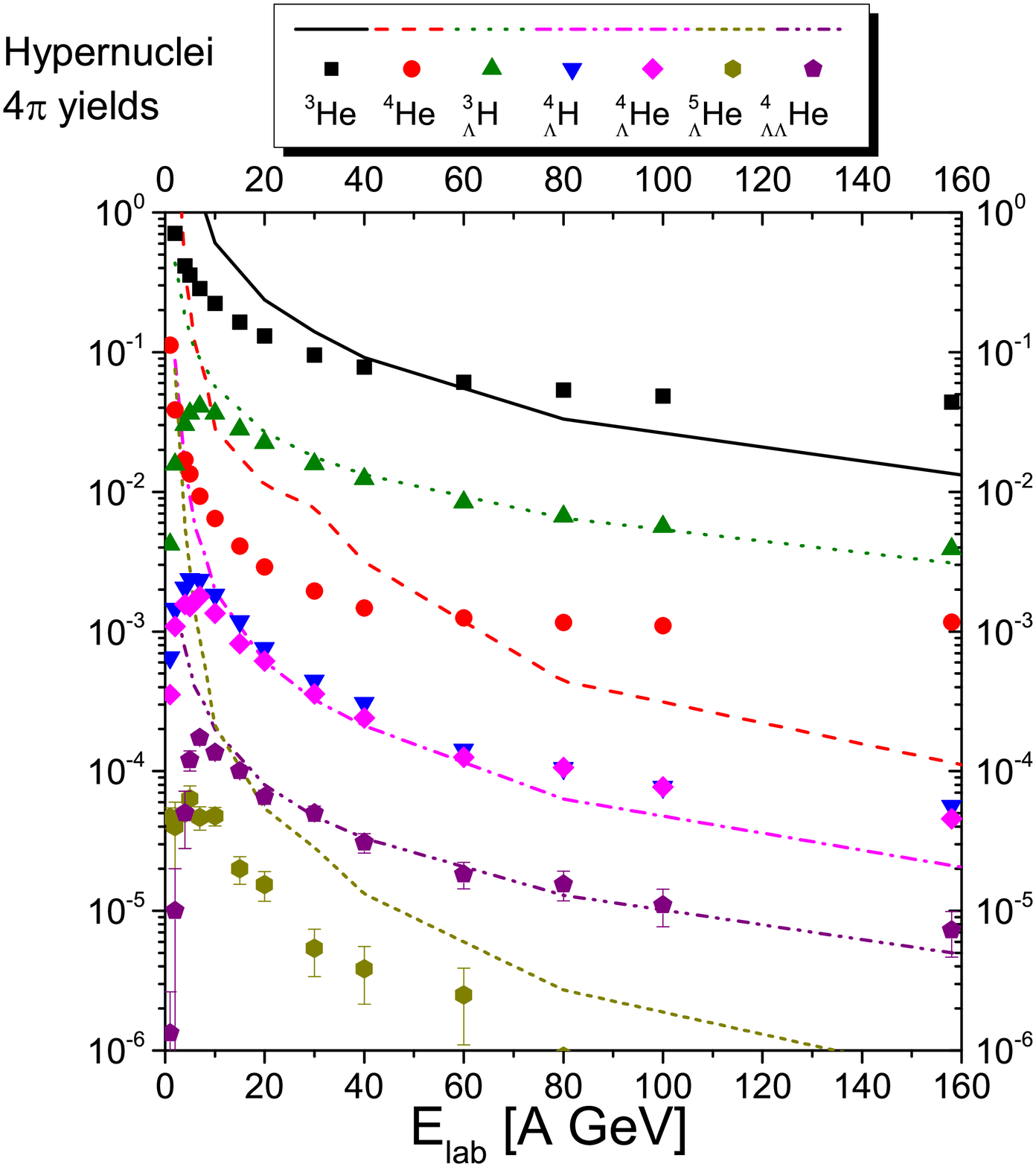}
\caption{Full acceptance yields per event of different (hyper-)nuclei created in most central collisions of Pb+Pb/Au+Au. Shown are the results from the thermal production in the UrQMD hybrid model (lines) as compared to coalescence results with the DCM model (symbols).  
\label{hypally}
}
\end{center}
\end{figure}

An instructive result is that the yields of most hypernuclei have a 
maximum (or saturation) around 10--20 A GeV of beam energy. Therefore, 
the investigation of hypernuclei can be effectively pursued at these energies. 
On the other hand, the dependence of their yields up to energies of $\sim$200 
A GeV can help to clarify the mechanisms of hypernuclei production. 

Noticeably the yields for di-baryons inlcuding $\Xi$ hyperons differ strongly with 
respect to the model applied, for the double $\Xi$ state the difference is as large 
as one order of magnitude. The reason for this discrepancy can be understood 
considering that the DCM model produces considerably, by a factor of 5 times, 
less $\Xi$'s than the UrQMD hybrid model, therefore also the dibaryon formation 
is strongly suppressed (note that the experimental $\Xi$ yield is quite well 
reproduced by the UrQMD-hybrid model \cite{Steinheimer:2009zzb,Steinheimer:2011mp}).\\

Di-baryon production rates have also been calculated in a coalescence approach using 
the RQMD model for $\sqrt{s_{NN}}=200$ GeV collisions of Au nuclei \cite{SchaffnerBielich:1999sy}. To relate 
our calculations to these results, they are indicated as the colored bars on the 
right axis of figure \ref{dibmidy}. The RQMD model used was in particular tuned 
to reproduce multi strange particle yields (such as the $\Xi$) and the results 
are therefore close to the ones obtained with our thermal/hydrodynamic approach. 
    
Figures \ref{dibally} and \ref{hypally} show the integrated ($4 \pi$) 
yields for all considered clusters as a function of beam energy. As with 
the midrapidity results there is a remarkable agreement between both approaches. 
However, the integrated yields of non-strange nuclei at high energies are 
systematically larger in the coalescence approach, although the mid-rapidity 
yield was smaller. This observation can be explained when the rapidity 
distribution of the nuclei is considered. In the coalescence approach the 
probability to produce a nucleus increases with rapidity and in particular 
in the fragmentation region, where the nucleons have small relative transverse 
momenta and can easily coalesce.

\begin{table}[b]
		\begin{tabular}{|c|c|c|c|c|c|}
		\hline 
$p_C$=& 5 & 20  & 50  & 90  \\ \hline\hline
  $\Lambda$N& 4.4 $\cdot 10^{-4}$  & 2.7 $\cdot 10^{-2}$ & 3.0 $\cdot 10^{-1}$& 2.1 \\ \hline
  $\Lambda\Lambda$& 3.0$\cdot 10^{-5}$ & 1.2$\cdot 10^{-3}$ & 6.6$\cdot 10^{-3}$ & 5.6$\cdot 10^{-2}$ \\ \hline
  $\Xi$N & $ < 10^{-6}$ & 1.0$\cdot 10^{-3}$ & 1.1$\cdot 10^{-2}$ & 1.0$\cdot 10^{-1}$  \\ \hline
  $\Xi\Lambda$ & $ < 10^{-6}$ & 7.4$\cdot 10^{-5}$ & 5.8$\cdot 10^{-4}$ & 1.0 $\cdot 10^{-2}$ \\ \hline
  $\Xi\Xi$ &   $ < 10^{-6}$ & $ < 10^{-6}$ & 3.8$\cdot 10^{-4}$  & 7.2$\cdot 10^{-4}$\\ \hline
  		\end{tabular}
	\caption{Dependence of yield of strange dibaryons (per one event) on momentum 
coalescence parameter ($p_C$ in units of [MeV/c]), in central $(b<3.5fm)$ Au+Au collisions 
at $20 A$ GeV \label{tablecoal}}
\end{table}

In addition we point out that the coalescence results depend 
on the parameters of the model. As mentioned, in the presented results the parameter $p_C$ for $\Lambda$'s was taken equal 
to the one of the nucleon's. However, the hyperon-hyperon 
and hyperon-nucleon interactions are not very well known 
and we expect that these parameters may be different for clusters containing 
$\Lambda$'s or even $\Xi$'s. In table \ref{tablecoal} we demonstrate how the yields of strange dibaryon nuclei 
depend on the momentum parameter $p_C$. As discussed previously, we have 
accordingly restricted the $r_C$ parameter, however, by imposing an 
empirical limitation related to the nuclear force properties that $r_C$ can 
not be larger than 4 fm. One can see, we expect a very large variation of 
the yields depending on the parameters. For instance, the probability of a bound 
$\Lambda$--nucleon state may decrease by many orders, if we assume a small $p_C$ 
corresponding to a low binding energy of this state. 
Usually the parameters are fixed by comparison 
with experiment. Nevertheless, ratios of hypernuclei yields should not be changed in 
the coalescence model. 

\begin{figure}[t]
\begin{center}
\includegraphics[width=0.5\textwidth]{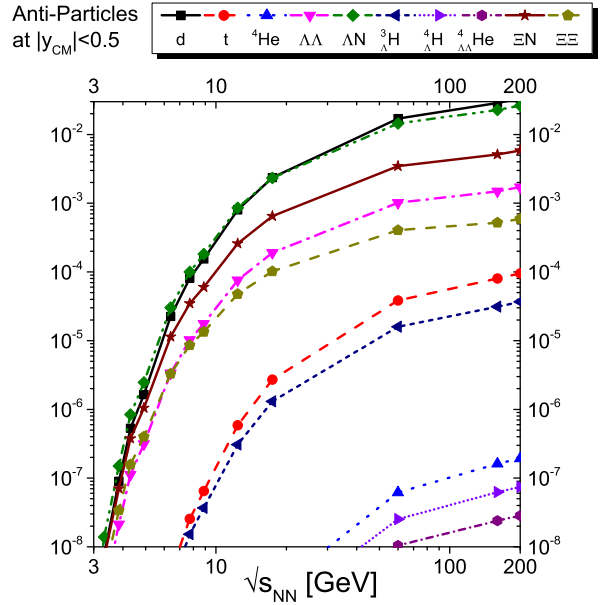}
\caption{Yields of anti-particle clusters in the mid rapidity region ($|y|<0.5$) of most central collisions of Pb+Pb/Au+Au as a function of $\sqrt{s_{NN}}$. Shown are only the results from the thermal production in the UrQMD hybrid model (lines with symbols). 
\label{antimidy}
}
\end{center}
\end{figure}

When the beam energy of the collisions is increased, the system created becomes 
almost net-baryon free. This means that the probability to create an anti-particle 
cluster approaches that of the particle cluster. Figure \ref{antimidy} shows the 
results for anti-particle cluster production at mid-rapidity ($|y|<0.5$) in 
collisions of Pb+Pb/Au+Au at center of mass energies of 
$\sqrt{s_{NN}}=3$ - $200$ GeV. We show only results for the UrQMD hybrid model 
because the DCM calculations are restricted to energies up to $E_{lab}=160 A$ GeV 
where the statistics needed for a meaningful estimate are quite significant. 
The yields of the anti-particle clusters show a monotonous increase with beam 
energy. They show that, at the highest RHIC energy (and at the LHC) the 
reconstruction of $_{\Lambda}^{4}\rm{He}$ might be a feasible task.


\subsection{A special ratio}
In the following we will discuss the double ratio $R_{H}$ defined as:
\begin{equation}
	R_{H}=_{\Lambda}^3 H / ^3 He \ \cdot p/\Lambda
\end{equation}
for collisions of Pb+Pb/Au+Au and a wide range of beam energies. This ratio is 
especially interesting, as in thermal production, it does not depend on the 
chemical potential of the particles (as fugacities cancel), and any canonical 
correction factors for strangeness are canceled. It has been proposed that this 
ratio is sensitive to the local correlation of strangeness and baryon number, 
therefore being a measure of $c_{BS}$ \cite{Zhang:2009ba}.\\

\begin{equation}
	c_{BS}=-3\frac{\left\langle N_B N_S\right\rangle- \left\langle N_B \right\rangle \left\langle N_S\right\rangle}{\left\langle N_S^2 \right\rangle - \left\langle N_S \right\rangle^2}
\end{equation}

To calculate $R_H$ we use the above obtained yields for hypernuclei and the 
proton and $\Lambda$ yields from the same model. For the hadrons the feed 
down from resonances is taken into account as well as the feed down to the 
$ ^3 He$ from the hypertriton.\\

\begin{figure}[t]
\begin{center}
\includegraphics[width=0.5\textwidth]{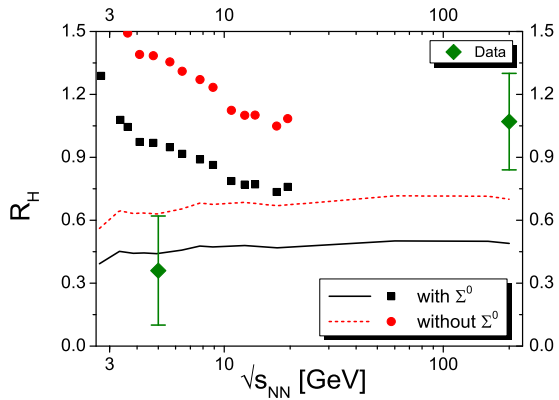}
\caption{The Strangeness Population Factor $R_H=(^{3}_{\Lambda}H/^{3}He)\cdot(p/\Lambda)$ as a function of $\sqrt{s_{NN}}$ for most central collisions of Pb+Pb/Au+Au. We compare results from the thermal production in the UrQMD hybrid model (lines) with coalescence results with the DCM model (symbols). The red line and symbols denote values of $R_H$ where the $\Lambda$ yield has been corrected for the $\Sigma^0$ contribution.
\label{rh}
}
\end{center}
\end{figure}

Our results for $R_H$ are shown in figure \ref{rh} as an excitation function 
of the beam energy $\sqrt{s_{NN}}$. $R_H$ is evaluated for the mid rapidity 
region of most central ($b<3.4$ fm) heavy ion collisions. The lines depict 
results from the UrQMD-hybrid model and the symbols denote DCM coalescence results. 
Experimental data are depicted as green symbols with error bars. Because 
experiments usually cannot distinguish between $\Lambda$'s and $\Sigma^0$'s, 
we show $R_H$ in the cases where the $\Lambda$ yield includes $\Sigma^0$ 
(black solid line and squares) and where the yield is corrected for the 
$\Sigma^0$ (red dashed line and circles). This is in fact important as there is 
no experimental indication for a bound $_{\Sigma^0}^3 H$ hypernucleus.\\

The double ratio $R_H$ from the hybrid model turns out to be almost energy 
independent. The same behavior has been observed in previous thermal calculations \cite{Andronic:2010qu}. 
On the other hand, the coalescence result increases with decreasing beam energy 
and is in general larger than the thermal result.

\begin{figure}[t]
\begin{center}
\includegraphics[width=0.5\textwidth]{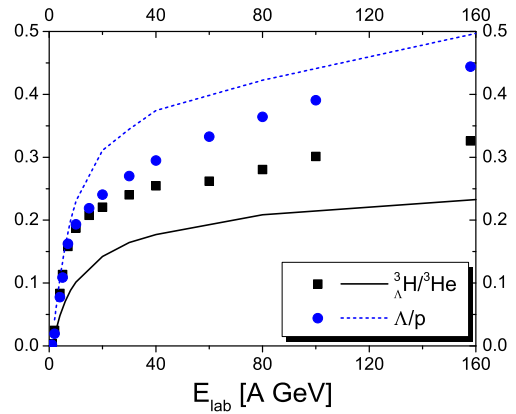}
\caption{Single ratios of $(^{3}_{\Lambda}H/^{3}He)$ (black solid line and circles) and $\Lambda/p$ (blue dashed line and squares) from the UrQMD hybrid model (lines) and DCM model (symbols).  
\label{special}
}
\end{center}
\end{figure}

To understand this behavior we plotted the single ratios 
$^{3}_{\Lambda}H/^{3}He$ and $\Lambda/p$ from our two approaches 
(lines hybrid model and symbols DCM coalescence) in figure \ref{special}. 
Here it is obvious that even though the DCM calculation produces less 
$\Lambda$'s per proton, the hypernuclei to nuclei ratio is still larger. 
Hence, the $\Lambda$ is more likely to form a hypernucleus. There seems to 
be a stronger correlation in the transport calculation as in the hydrodynamic 
description. In fact the qualitative behavior of $R_H$ closely resembles the 
behavior that is expected for $c_{BS}$, the baryon-strangeness correlation, 
for a hadronic gas \cite{Koch:2005vg}. This observation leads to the conclusion that the 
information on correlations of baryon number and strangeness is lost in the 
thermal calculation because here $R_H$ essentially only depends on the 
temperature. On the other hand, in the microscopic treatment the correlation 
information survives and $R_H$ captures the trend of $c_{BS}$.

\section{Conclusion}

We have presented results on hyper-nuclei, anti-nuclei and di-baryon production
in heavy ion collisions over a wide beam energy range. To explore the
theoretical uncertainties we applied two distinct approaches: firstly, the
thermal production with the UrQMD-hydro hybrid model and secondly, the
coalescence calculation within the Dubna hadron cascade model. Concerning most
hyper-nuclei and di-baryons both approaches agree well in their predictions
which gives us confidence in robustness and significance of the obtained
results. We find that both the non-equilibrium and thermal models may be
considered as appropriate approaches to describe strange cluster
production. In agreement with previous studies we demonstrate that the most
promising energy range to produce hyper-clusters will be provided by the FAIR
and NICA facilities, $E_{lab}\approx 10$ - $20 A$ GeV. Anti-matter clusters
heavier than $\bar{t}$ are only feasible at RHIC and LHC energies.\\

The most interesting result of our study is the apparent difference in the
double ratio $R_H$ when we compare our thermal results with the coalescence.
This difference indicates that the information on correlations of baryon number
and strangeness are visible in the microscopic coalescence approach, while they
are washed out in the thermal picture. This could open the opportunity to
directly measure the strangeness-baryon correlation, which may be sensitive to
the onset of deconfinement. The present status of the experimental data does
unfortunately not allow for a comprehensive comparison with our model
calculations. We hope that this situation will improve in the upcoming RHIC
energy scan and FAIR experiments.
\section*{Acknowledgments}

This work has been supported by GSI and Hessian initiative for excellence (LOEWE) 
through the Helmholtz International Center for FAIR (HIC for FAIR). J.~S. acknowledges a Feodor
Lynen fellowship of the Alexander von Humboldt foundation. This work was supported by the Office of Nuclear Physics in the US
Department of Energy's Office of Science under Contract No. DE-AC02-05CH11231. I.M. acknowledges partial support from grant NS-215.2012.2 (Russia). The computational resources were provided by the LOEWE Frankfurt Center for Scientific Computing (LOEWE-CSC).

\end{document}